\documentclass[showpacs,twocolumn,prl,preprintnumbers,amsmath,amssymb,showkeys]{revtex4}

\usepackage{graphicx}
\usepackage{dcolumn}
\usepackage{color}
\usepackage{subfigure}
\usepackage{bm}
\usepackage{bezier}

\begin{document}

\title{General relaxation time of the fidelity for isolated quantum thermodynamic systems}
\author{Takaaki Monnai}
\email{monnai@suou.waseda.jp}
\affiliation{Waseda Institute for Advanced Study, Waseda University, Tokyo 169-8050, Japan}
%

\date{\today}

\begin{abstract}
General evaluation of the relaxation time to equilibrium is usually considered as difficult, since it would strongly depend on the model of interest.
In this paper, we provide a generic initial relaxation time of the fidelity for the isolated large systems. The decay of the fidelity is a combination of the Lorentzian and a sinusoidal oscillation. 
We calculate the relaxation time of the Lorentzian envelop, and the period of the oscillation. Remarkably, these two time scales are the same order when the energy range of the microcanonical state is larger than the thermal fluctuation. 
Also, the power law decay generally exists for long time regime.                         
\end{abstract}
\maketitle
\section{Introduction}
Recently, considerable attentions have been paid to the thermalization of quantum many-body isolated systems. The central problems of the thermalization are to reveal the mechanism of the relaxation to equilibrium, and the general evaluation of the relaxation time.
Regarding the mechanism of the relaxation, the fact that for a limited number of observables, typical pure states yield expectation values very close to thermal average provides an important progress\cite{Lebowitz1,Popesucu,Sugita,Reimann,Rigol}. In particular, it is important that only a single typical pure state is enough to analyze equilibrium\cite{Sugiura1} and nonequilibrium processes\cite{Monnai1}. Relaxation and recurrence dynamics of the Lieb-Liniger model were studied as well\cite{Deguchi1,Deguchi2}. For these solvable cases, the commensurability of the energy spectrum amounts to remarkably short recurrence time. For generic systems, the system size dependence of the recurrence time was reported in Ref. \cite{Peres}, which shows that the recurrence time hyper exponentially depends on the system size.   
On the other hand, only a few is known for the evaluation of the relaxation time.
Ref. \cite{Tasaki1} rigorously shows a possibility of an extremely slow decay for some initial states, and fast decay occurs for randomly chosen initial states\cite{Tasaki2}.  
In Ref. \cite{Monnai2}, the relaxation time of expectation values are theoretically evaluated based on several assumptions i-iii) such as i) nonintegrability, ii) preparation of the initial nonequilibrium, and iii) monotonic approach to the equilibrium.  

For open systems in contact to a large reservoir, the fidelity has been calculated as a standard measure of relaxation. Indeed, it revealed the presence of three nontrivial time scales, i.e., quantum Zeno regime\cite{Misra1}, Wigner-Weisskopf exponential decay, and power law decay\cite{Fonda1}. The fidelity has been also used to characterize how the small perturbation affects the time evolution\cite{Weinstein1}.

The purpose of this paper is to investigate the relaxation dynamics of the fidelity for isolated thermodynamic systems. 
In this way, we have a general relaxation time of the fidelity based on the general thermodynamic property of the density of the states. In particular, we show that the relaxation consists of Lorentzian decay and sinusoidal oscillation. The relaxation time in the present paper is compatible to Refs. \cite{Tasaki2,Monnai2}. 
It provides a first step to analyze the generic relaxation time.

This paper is organized as follows.
In Sec. 2, we establish our general model.
In Sec. 3, we calculate the fidelity for thermodynamic parameter regimes.
In Sec. 4, the fidelity is numerically calculated for a spin chain. 
Sec. 5 is devoted to a summary.
\section{Model}
We consider a many-body isolated quantum system, whose energy scale $E$ is determined with a precision $\Delta E$.   
The Hilbert space ${\cal H}_{[E,E+\Delta E]}$ is spanned by the eigenenergy states $\{|E_n\rangle\}$ $(1\leq n\leq d)$, where $d={\rm dim}{\cal H}_{[E,E+\Delta E]}$ is the dimension.
Suppose that the initial state is expanded in the eigenenergy basis as
\begin{equation}
|\phi(0)\rangle=\sum_{n=1}^d c_n|E_n\rangle.
\end{equation}
The expansion coefficients $c_n=|c_n|e^{i\phi_n}$ satisfies the normalization condition
\begin{equation}
\sum_{n=1}^d|c_n|^2=1.
\end{equation}
For simplicity of our analysis, we assume that amplitudes of all the coefficients are exactly the same $|c_n|^2=\frac{1}{d}$.
To describe the relaxation, we are usually interested in a set of observables $\{A_n\}$, which consists a small subset of all the Hermitian operators and significantly deviate from equilibrium value at initial time. Here, for concreteness, we consider the case that an observable $A$ in $\{A_n\}$ shows significant deviation from equilibrium.       
Indeed, we can successfully prepare the nonequilibrium initial state by properly choosing the phases $\{\phi_n\}$ so that the expectation value of an observable $A$ at $t=0$ significantly deviates from its equilibrium value. It means that  the initial state $|\phi(0)\rangle$ is not a typical state whose expectation value of $A$ gives microcanonical average with probability very close to unity. And, we investigate how the relaxation of such an initial nonequilibrium state occurs.

Before calculating the fidelity, we want to make clear the relevance of its use. The fidelity reveals how the initial phase relation disappears in the course of time evolution. 
Our numerical simulation strongly suggests that the fidelity is $O(\frac{1}{d})$ after the relaxation, and the initial phase coherence completely disappears. It is then reasonable to expect that initial nonequilibrium state evolves to another state within the relaxation time of the fidelity, which is usually regarded as equilibrium, since the majority of the pure states are equilibrium.
Therefore, we regard the above-mentioned observable $A$ also relax to equilibrium. Then, it is advantageous to use the fidelity, because we can analytically evaluate the relaxation time. On the other hand, the analysis of the fidelity alone would not be sufficient to fully characterize rich variety of relaxation processes. A drawback in this way is that even when the fidelity is very small, there is still a possibility that states at different times can give macroscopically the same expectation values for some quantities of interest $A_n(\neq A)$. Preparation of an initial nonequilibrium state for all the observables in $\{A_n\}$ provides an interesting future problem.
                  
One may think that the fidelity would be too strict as a measure of the distance. Indeed, it is unstable against a local perturbation when product states are concerned. However, we are interested in a superposition of them, which is usually entangled. For this reason, the inner product of superposition states gradually decays as we will show later.     
         
The inner product of the wave functions at time $0$ and $t$ is thus
\begin{eqnarray}
&&\langle\phi(0)|\phi(t)\rangle=\sum_{n=1}^d|c_n|^2e^{-\frac{i}{\hbar}E_n t} \nonumber \\
&=&\frac{1}{d}\sum_{n=1}^de^{-\frac{i}{\hbar}E_n t}. \label{inner1}
\end{eqnarray}
\section{Thermodynamic systems}
Suppose that the system size $N$ is large, and the dimension is exponentially large $d=O(e^N)$.    
Then, we can take a continuous limit, and Eq. (\ref{inner1}) is rewritten as
\begin{eqnarray}
&&\langle\phi(0)|\phi(t)\rangle \nonumber \\
&\cong&\frac{1}{\Delta E}\frac{\Delta E}{d}\sum_{n=1}^d e^{-\frac{i E}{\hbar}t}e^{\frac{i n\Delta E}{\hbar d} t}\Omega(E+\frac{n\Delta E}{d}) \nonumber \\
&\cong &\frac{1}{\Delta E}e^{-\frac{i E}{\hbar}t}\int_0^{\Delta E}dx e^{-\frac{i x}{\hbar}t}\Omega(E+x), \label{fidelity1}
\end{eqnarray}
where we defined a sort of density of the states $\Omega(E+\frac{n\Delta E}{d})$.
More precisely, $\Omega(E+\frac{n\Delta E}{d})$ is the number of the eigenstates which satisfy $E+\frac{n\Delta E}{d}\leq E_m\leq E+\frac{(n+1)\Delta E}{d}$.
Thus, we have $\sum_{n=1}^d\Omega(E+\frac{n\Delta E}{d})=d$. 
In the continuous limit, $\frac{1}{\Delta E}\Omega(E+x)dx$ is the number of the eigenstates which satisfy $E+x\leq E_n\leq E+x+dx$. 
Note that $\Omega(E+x)$ is dimensionless, however, we simply call this quantity as density of the states. Indeed, $\Omega(E)$ is proportional to the density of the states. 

Here is a remark on the evaluation of Eq. (\ref{fidelity1}). The density of the states rapidly increases as the energy gets larger.
Then, we need to consider two cases i) $\beta\Delta E> 1$ and ii) $\beta\Delta E\leq 1$. Here,  $\beta$ is the inverse temperature. 
The choice of $\Delta E$ is in principle independent from the temperature.
However, in the thermodynamic systems, we usually consider the former case, and our main interest lies in the case i).
We consider the case ii) to compare the analytic calculation with quantum mechanical numerical simulations\cite{Shankar}, as we will show later.
 
i) If the energy range $\Delta E$ is much lager than the thermal fluctuation, $\Omega(E+x)$ is dominant only for $x$ in $[\Delta E_0,\Delta E]$, where the cutoff $\Delta E_0$ satisfies  $\beta(\Delta E-\Delta E_0)=O(1)$. ii) On the other hand, when $\Delta E$ is smaller than the thermal fluctuation, the cutoff is zero $\Delta E_0=0$. Hereafter, we abbreviate $\Delta E-\Delta E_0$ as $\Delta E_{\rm eff}$.

Thus, we can expand the density of the states as 
\begin{equation}
\log\Omega(E+\Delta E_0+x)=\log\Omega(E+\Delta E_0)+\beta x+O(x^2) \label{entropy1}
\end{equation}    
up to the first order for thermodynamic systems.
It is remarked that we can identify $\beta=\frac{\partial}{\partial E}\log\Omega(E)|_{E+\Delta E_0}$ as the inverse temperature.
Here, we set the Boltzmann constant unity $k_B=1$.
The second order is $-\frac{\beta^2}{2C_V}x^2$, 
where the heat capacitance $C_V$ is proportional to the system size.
The second order is negligible compared with the first order when $C_V\gg \beta\Delta E_{\rm eff}$.
This condition is satisfied for the thermodynamic systems, since the heat capacitance is large, and $\beta\Delta E_{\rm eff}=O(1)$. 

Thus, the inner product is further calculated as
\begin{eqnarray}
&&\langle\phi(0)|\phi(t)\rangle \nonumber \\
&\cong&\frac{1}{\Delta E}e^{-\frac{i E}{\hbar}t}\int_{\Delta E_0}^{\Delta E}dx e^{-\frac{i x}{\hbar}t+\log\Omega(E+\Delta E_0)+\beta x} \nonumber \\
&=&\frac{\Omega(E+\Delta E_0)}{\Delta E}e^{-\frac{i (E+\Delta E_0)}{\hbar}t}e^{\beta\Delta E_0} \nonumber \\
&&\times\frac{1}{-\frac{i}{\hbar}t+\beta}(e^{-\frac{i\Delta E_{\rm eff}}{\hbar}t}e^{\beta \Delta E_{\rm eff}}-1).
\end{eqnarray}
Therefore, the square of the absolute value gives 
\begin{eqnarray}
&&F(t)\cong\frac{\Omega(E+\Delta E_0)^2}{\Delta E^2} \nonumber \\
&&\times\frac{1}{\frac{t^2}{\hbar^2}+\beta^2}(1+e^{2\beta\Delta E_{\rm eff}}-2 e^{\beta\Delta E_{\rm eff}}\cos\frac{\Delta E_{\rm eff}}{\hbar}t). \label{probability1}
\end{eqnarray}

Importantly, the fidelity is a product of the Lorentzian and the oscillatory term.
There are thus two time scales, i.e.,  the relaxation time for the Lorentzian $T_1=\beta\hbar$ and the period of the oscillation $T_2=\frac{2\pi\hbar}{\Delta E_{\rm eff}}$. i) For thermodynamic systems, we have $\beta\Delta E_{\rm eff}=O(1)$ and these two time scales coincide $T_1\cong T_2$. 
ii) On the other hand, for the numerical cases $\beta\Delta E\ll 1$, the period of the oscillation is much longer than the relaxation time of the Lorentzian $T_1\ll T_2$. 
The fidelity at $t=T_2$ is $\frac{\Omega(E)^2}{(2\pi)^2+(\beta\Delta E)^2}(1-e^{\beta\Delta E})^2\ll 1$ for $\beta\Delta E\ll 1$.
Therefore, we regard $T_2=\frac{2\pi\hbar}{\Delta E_{\rm eff}}$ as the relaxation time.
Another important point is that the fidelity shows power law decay for long time regime.
Interestingly, this is compatible with the power law decay reported for some solvable models in the presence of the infinitely large reservoir except for the exponent 
and Pailey-Wiener's theorem for Fourier-Laplace transformation\cite{Fonda1}. On the other hand, the Wigner-Weisskopf  exponential decay due to the interaction with the reservoir is absent in our case. The decay rate is usually given by the strength of the interaction, however, $T_2$ only depends on $\Delta E_{\rm eff}$, which is determined by the initial condition.    
Having recourse to the isolated systems, we could show the initial relaxation dynamics and  the slow decay for generic large systems.  
It would mean that the concept of the relaxation time is well-defined only for the short time.   
\section{Numerical simulation}
In this section, we analyze the relaxation phenomenon by the numerical simulation for the case ii) $\beta\Delta E\ll 1$.
We consider a spin-chain in a magnetic field\cite{Monnai2,Shankar}, since it shows thermodynamic nature for relatively small system size $N\cong 7$.
The Hamiltonian 
\begin{equation}
H=-J\sum_{j=1}^{N-1}\sigma_j^z\sigma_{j+1}^z+\alpha\sum_{j=1}^N\sigma_j^x+\gamma\sum_{n=1}^N\sigma_j^z
\end{equation}
consists of the nearest neighbor coupling and the magnetization.
\begin{figure}
\center{
\includegraphics[scale=0.6]{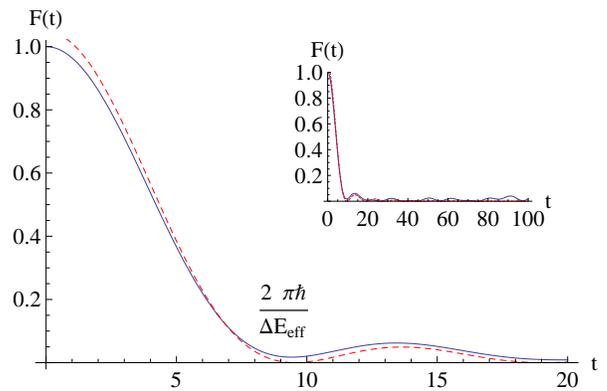}
}
\caption{The time evolution of $F(t)$ (blue-line). The theoretical value for long time regime Eq. (\ref{probability1})(red broken-line) well-agrees with the numerical data after the initial transient. The parameters are $\beta=0.07$ and $\Delta E=0.658$, which yield  $T_1=0.046$ and $T_2=9.55$. Inset shows that $F(t)$ remains $O(10^{-2})$ due to the finite system size.}
\end{figure}
We choose the parameters as $J=1$, $\alpha=1$, and $\gamma=0.5$.
We note that $\gamma\neq 0$ corresponds to the nonintegrable case.
The system size is $N=10$.
Then, the Hamiltonian has $2^N$ eigenenergies.
For example, we consider the Hilbert space ${\cal H}_{[E,E+\Delta E]}$ spanned by the $201$-th to $240$-th excited states, where the inverse temperature is $\beta=0.07$ and $\Delta E=0.658$, which actually satisfies the condition $\beta\Delta E\ll 1$. And, $\Delta E_{\rm eff}=\Delta E$. The energy range is small due to the relatively small system size.   The density of the states $\Omega(E+x)$ is almost constant for most $E\leq x\leq E+\Delta E$. 
We randomly choose an initial state $|\phi(0)\rangle=\sum_{n=201}^{240} c_n|E_n\rangle$ from ${\cal H}_{[E,E+\Delta E]}$. We set the Planck constant unity $\hbar=1$.
The coefficients are chosen from complex valued 
 random variables, and the mean of the $|c_n|^2$ is $0.025$ and the variance is $0.000177$. 
The minimum and maximum values of the square of absolute values  of coefficients $|c_n|^2$ are $0.00089$ and $0.047$. 
Thus, the amplitude of the coefficients are distributed around the mean value. And, the variance of the phase in mod $2\pi$ is $0.173$. 
In Fig. 1, we compare the numerical fidelity and Eq. (\ref{probability1}).
We have a good agreement for $3\leq t\leq 40$.  
For $t\leq 3$, the fidelity shows parabolic behavior as the consequence of the unitary evolution. Then, the fidelity almost linearly decreases until $t=6$.    
And, the relaxation time is actually given by $T_2$.  
For $t\geq 40$, the numerical fidelity is $O(10^{-2})$ and does not converge to zero due to the finite system size as shown in the inset of Fig. 1.
It would mean that $|\phi(0)\rangle$ and $|\phi(t)\rangle$ are totally uncorrelated, and the fidelity is of order $\frac{1}{d}=0.025$.  
It is also possible to numerically take into account the second order contribution in Eq. (\ref{entropy1}). In this way, we can take into account the convexity of the entropy. However, the deviation from Eq. (\ref{probability1}) is very small, whose mean variance during $0\leq t\leq 20$ is $2.73\times 10^{-6}$. And, the truncation in Eq. (\ref{entropy1}) is reasonable.

If the initial state $|\phi\rangle$ is a product state $|+,...,+\rangle$, then the fidelity would soon decay. In order to make clear this point, we have calculated   $|\langle\sigma_1,...,\sigma_{N}|e^{-iH t}|+,...,+\rangle|$ and $|\langle +,...,+|e^{-iH t}|+,...,+\rangle|^2$. Here, $|\sigma_j\rangle=|\pm\rangle$ are the eigenstates of $\sigma_j^z$. First, it is remarked that $|+,+,...,+,+\rangle$ is composed of eigenenergy states broadly distributed from the $21$-th excited state to $837$-th excited states, and the energy scale is not well-defined. For $t=2$, the mean and variance of such quantities are $0.0249$ and $0.000356$. The diagonal element is also small  $|\langle +,...,+|e^{-iH t}|+,...,+\rangle|=0.116$. The fidelity quickly relaxes until $t=1$, which is understood that $\Delta E=12.6$ is large compared with the range of the energy shell.
The relaxation time monotonically decreases as a function of the strength of the nearest neighbor interaction $J$. For $J=0.5, 1, 2, 5$, we have $T_2=0.86, 0.67, 0.51, 0.42$, respectively.
Thus, the initial product state is completely destroyed within a time much shorter than the relaxation time $T_2$.     

For further insights, we also plot the normalized theoretical values Eq. (\ref{probability1}) for various $\Delta E_{\rm eff}$ with fixed $\beta\Delta E_{\rm eff}=1$ in Fig. 2. It provides numerical evidences that the relaxation time is given by $T_2$, i.e. $2\pi T_1$.

\begin{figure}
\center{
\includegraphics[scale=0.8]{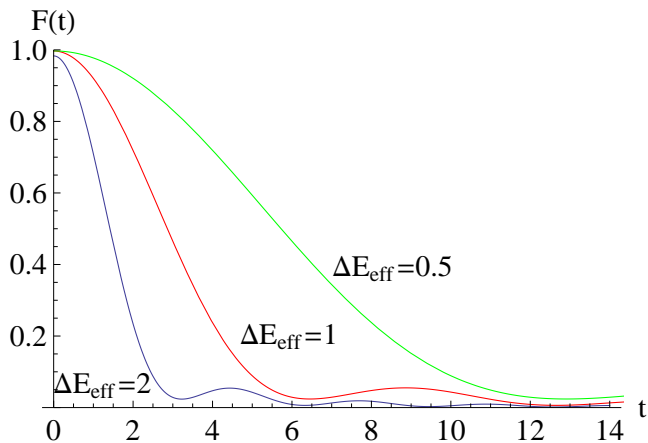}
}
\caption{The time evolution of $F(t)$ calculated by Eq. (\ref{probability1}) for $(\Delta E_{\rm eff},\beta)=(2,0.5)$ (blue-line), $(1,1)$ (red-line), and $(0.5,2)$ (green-line). The two time scales $(T_1,T_2)$ are $(0.5,\pi)$, $(1,2\pi)$, and $(2,4\pi)$, respectively.}
\end{figure}      
\section{Summary}
We analytically calculated 
 the temporal behavior of $F(t)$ for isolated large systems.
The relevance to use the fidelity in the analysis of relaxation phenomenon is discussed in detail. In particular, we derived the relaxation time $T_1$ of the Lorentzian envelop and period of oscillation $T_2$ for the fidelity. We have numerically confirmed that $F(t)$ is $O(\frac{1}{d})$ after $T_2$ for the case ii) $\beta\Delta E\leq 1$.  Eq. (\ref{probability1}) holds as well for the case i) $\beta\Delta E>1$, since the correctness of the second order expansion of the entropy is well-established for thermodynamic systems. Then, the initial relaxation time is given by $T_2=\frac{2\pi\hbar}{\Delta E_{\rm eff}}$. In particular, for the experimentally accessible case i) $\beta\Delta E\gg 1$, the relaxation time is the same order as the so-called Boltzmann time $T_1=\beta\hbar$, which is compatible to Refs. \cite{Tasaki2,Monnai2}.
Further analysis of relaxation time demands careful choice of the observables and Hamiltonian, and remains as no-man's land.


\section{Acknowledgment}
The author is grateful to Professor K. Yuasa, Professor H. Tasaki, and Professor T. Deguchi for fruitful discussions. 
This work is financially supported by Waseda University Grants for Special Projects( 2013A-982).

\end{document}